# Single-Atom Gating of Quantum State Superpositions

Christopher R. Moon[1], Christopher P. Lutz[2] & Hari C. Manoharan[1*]

[1]*Department of Physics, Stanford University, Stanford, California 94305, USA*
[2]*IBM Almaden Research Center, 650 Harry Road, San Jose, California 95120, USA*

*To whom correspondence should be addressed. E-mail: manoharan@stanford.edu

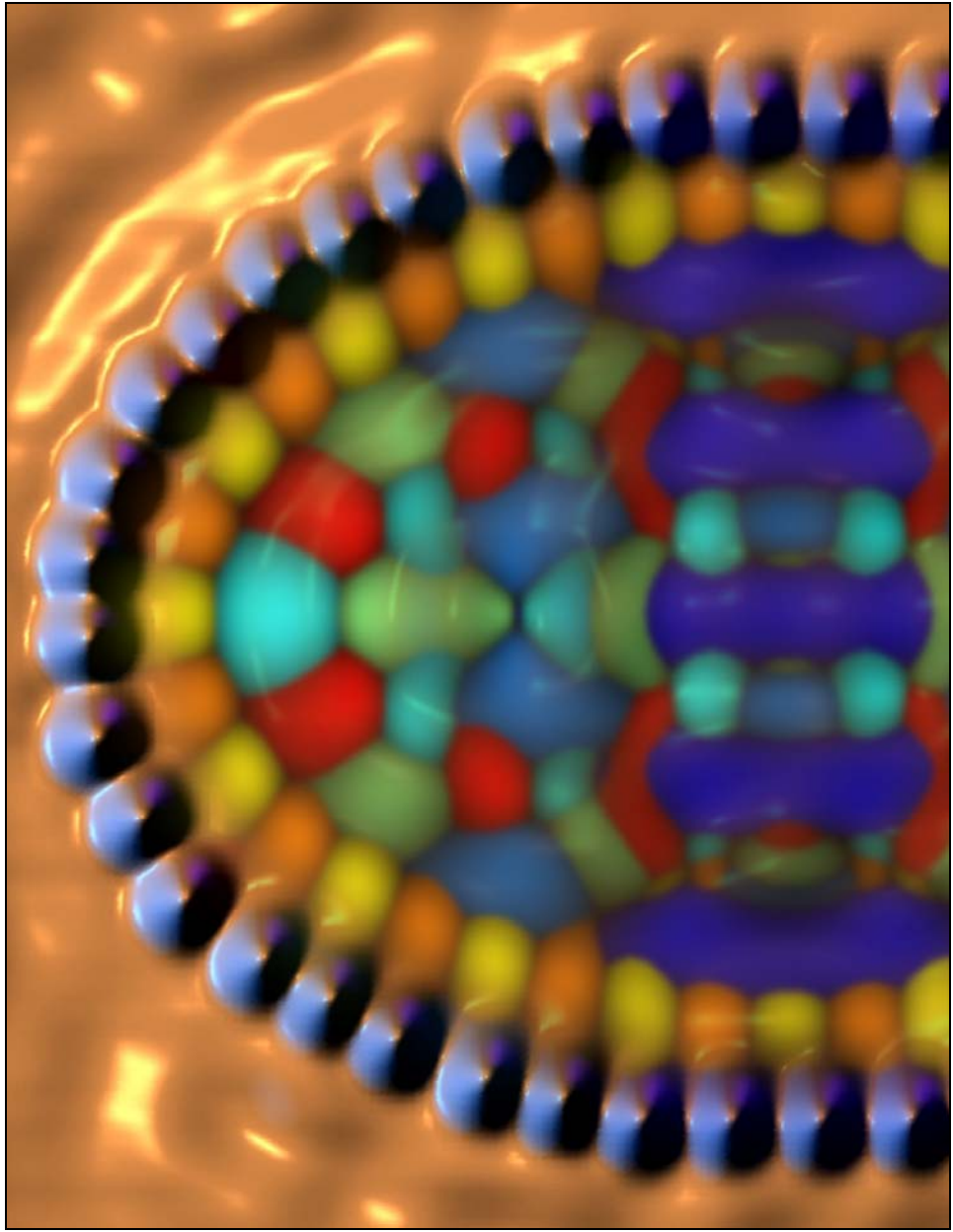

**The ultimate miniaturization of electronic devices will likely require local and coherent control of single electronic wavefunctions. Wavefunctions exist within both physical real space and an abstract state space with a simple geometric interpretation: this state space—or Hilbert space—is spanned by mutually orthogonal state vectors corresponding to the quantized degrees of freedom of the real-space system. Measurement of superpositions is akin to accessing the direction of a vector in Hilbert space, determining an angle of rotation equivalent to quantum phase. Here we show that an individual atom inside a designed quantum corral[1] can control this angle, producing arbitrary coherent superpositions of spatial quantum states. Using scanning tunnelling microscopy and nanostructures assembled atom-by-atom[2] we demonstrate how single spins and quantum mirages[3] can be harnessed to image the superposition of two electronic states. We also present a straightforward method to determine the atom path enacting phase rotations between any desired state vectors. A single atom thus becomes a real-space handle for an abstract Hilbert space, providing a simple technique for coherent quantum state manipulation at the spatial limit of condensed matter.**

Progress in quantum nanoscience has engendered a physically diverse array of controllable solid-state quantum systems[4-6]. The prototypical quantum system consists of two wavefunctions that can be coherently combined into superpositions. In this work, we create and study superpositions of electron wavefunctions in nanoassembled quantum corrals where we can finely tune the geometry. These structures permit unique investigations of nanoscale electrons and their correlations, including information propagation[7], lifetime effects[8], Kondo interactions[9], and spin-orbit coupling[10]. We use quantum corrals to take the traditional technique of gating, or the application of electrostatic potentials, to its smallest possible scale. Here, a single adatom couples only to a minute fraction of an electron's spatial extent. Rather than changing applied voltages, we make only geometric changes to the gate position, enabling adiabatic control of a two-state quantum system. Geometric and adiabatic manipulation of wavefunctions is a robust alternative to dynamic manipulation for various quantum technologies[11, 12].

We engineered an elliptical resonator to harbour degenerate wavefunctions whose superpositions could be manipulated. The solutions to the Schrödinger equation in a hard-walled ellipse possess two quantum numbers: $n$, the number of nodes crossing the minor axis, and $l$, half the number of nodal intersections along the perimeter (these map to the radial and angular momentum quantum numbers in a circle). By judiciously choosing an ellipse's deformation and size, two target wavefunctions can be made degenerate at any energy. We targeted states with even $n$ and non-zero $l$, which possess two widely separated regions of concentrated amplitude along the major axis. After analytically solving for the ellipse's eigenspectrum as a function of deformation (Fig. 1a), we aimed to design a corral with states $|n,l\rangle = |4,4\rangle$ and $|2,7\rangle$ degenerate precisely at the Fermi energy $E_F$. These correspond to the 41$^{st}$ and 42$^{nd}$ most energetic states, or $\psi_j \equiv |j\rangle = |41\rangle$ and $|42\rangle$. On Cu(111), where the surface state band edge is 0.445 eV below $E_F$ and $m^* = 0.38$ bare electron masses[13], this is theoretically achieved in a $2a \times 2b = 157 \times 110$ Å elliptical resonator, whose full energy spectrum is shown in Fig. 1b.

We assembled our designed resonator using a home-built scanning tunnelling microscope (STM) operating in ultrahigh vacuum. The single-crystal Cu(111) substrate was prepared, cooled to ~4 K, and dosed with ~15 Co atoms per (100 Å)$^2$. We individually manipulated[2] 44 Co adatoms to bound the corral. With spectroscopy, we verified that modes $|41\rangle$ and $|42\rangle$ occurred within a few mV of one another (Supp. Fig. 1). A constant-current ($I$) topograph of the finished structure is shown in Fig 1c. To confirm that the wavefunctions $|\psi\rangle$ closely describe this system, we used them to calculate (see Methods) a theoretical topograph (Fig. 1d) that reproduces the data without any fitting parameters. Figure 1e displays the calculated contributions $c_j$ of the significant modes composing the topograph $z(\mathbf{r})$, such that $z(\mathbf{r}) \propto \sum_j c_j \left\| \psi_j(\mathbf{r}) \right\|^2$.

Next, we added a nanoscopic gate: a single cobalt atom. While moving the adatom across the ellipse—effectively sweeping a local electrostatic potential across the eigenstates—we measured topographs (Fig. 2, first column) and simultaneously acquired $dI/dV$ image maps. By subtracting the $dI/dV$ map of the empty ellipse, we created $dI/dV$ difference maps (Fig. 2, second column). We began by placing the gate atom at one of the maxima of the calculated $|2,7\rangle$ state. The resultant difference map (Fig. 2e) strongly resembles the $|2,7\rangle$ state. Surprisingly, however, when the Co atom was moved rightward to one of the strong maxima of the state $|4,4\rangle$, the image produced (Fig. 2g) was manifestly different from either of the two eigenstates.

We will show that our $\Delta dI/dV$ maps are images of superpositions: phase-coherent $\left| \sum_j a_j |\psi_j\rangle \right|^2$. This is in contrast to typical STM measurements, such as the topograph above, where the sum of tunnelling through independent channels[14] yields signals proportional to phase-insensitive $\sum_j c_j \left\| \psi_j\rangle \right\|^2$. To demonstrate this result, we reproduce the difference maps as linear combinations of the states $|\psi\rangle$ of the unperturbed elliptical corral. These coherent superpositions (Fig. 2, third column) are an excellent match to the mirage data. Any methods neglecting phase interference cannot reproduce our observations (see Supp. Fig. 2).

Electrons in quantum corrals are well modelled by particle-in-a-box solutions to the Schrödinger equation because the surface state wavelength [30 Å in Cu(111) at $E_F$] is much larger than the spacing between the wall atoms[1, 3, 10, 15-18]. As a first clue to the underlying physics, the original report of the quantum mirage[3] pointed out the similarity between the solitary eigenfunction closest to $E_F$ and the spatial fine structure around the projected Kondo image. A complementary approach[10, 19-24] treats the wall atoms as discrete scatterers of electron waves and solves the quantum multiple-scattering problem. In fact, the scattering theory was also applied[20] to produce very compelling computed matches to the mirage data and extract the Kondo phase shift. In this prior experimental and theoretical work, the structures studied had modes separated in energy by more than their linewidths—i.e. the state manifold was essentially non-degenerate. Here we focus on the engineered degenerate case where two or more states are forced to coherently superpose by the added gate atom.

While the exact origin of the Kondo resonance in our system is still debated[25, 26], microscopic calculations have indicated that the many-body complexity is manifested in the energy dependence of the density of states while its spatial dependence can be understood from a single-particle perspective[16, 18]. Regardless of its origin, we have verified that Kondo scattering calculations[21] can reproduce many of the details of our images. However, the simplest and most intuitive model that reproduces the data requires only two electron wavefunctions and their elementary superpositions. Indeed, the success of our analysis is perhaps surprising in light of the well-known softness of the corral walls[19]. Without negating more complex approaches, this work vindicates the simple eigenmode picture of quantum corrals, which is relevant and useful to a variety of emerging applications.

For each superposition in Fig. 2, we selected the coefficients $a_j$ to create $\Psi(\mathbf{r};\mathbf{R}) = \sum_j a_j(\mathbf{R})|\psi_j(\mathbf{r})\rangle$, where $\mathbf{R}$ is the atom position and $\mathbf{r}$ a position within the ellipse, subject to the normalization constraint $\sum_j |a_j|^2 = 1$. If only two wavefunctions participate for a given $\mathbf{R}$, the state can be represented by a vector on a Bloch sphere: $|\Psi\rangle = \cos(\theta)|\psi_1\rangle + e^{i\phi}\sin(\theta)|\psi_2\rangle$. However, time-reversal symmetry requires that $|\Psi\rangle$ be real, so we are restricted to $\phi \in \{0, \pi\}$. Thus, we describe the mirage-projecting superposition with a single phase angle $\theta$, via $|\Psi\rangle = \cos(\theta)|\psi_1\rangle + \sin(\theta)|\psi_2\rangle$. Supplementary Video 1 shows how $|\Psi\rangle$ evolves as the phase angle is varied in the space composed of $|2,7\rangle$ and $|4,4\rangle$. The angles used to create the superpositions for each gate atom location are indicated in the fourth column of Fig. 2. This decomposition demonstrates that by translating the atom in real space, we are effecting a rotation in the Hilbert space spanned by the two unperturbed states. For example, nudging the atom back by 5 Å from the position in Fig. 2c to 2b achieves a state-space rotation of 30°.

The full Hilbert space of unperturbed wavefunctions is richer than the 2-space sampled in the first three rows of Figure 2. Seven modes—$|39\rangle$ through $|45\rangle$, the set shown in Fig. 1e—significantly overlap both the energy range of our tunnelling electrons and the energies where Kondo processes can occur (governed by the Kondo temperature $T_K$). However, at a given position **R** most of these modes have little amplitude and can be disregarded. For example, the atom positions in Fig. 2a-c lie within a subspace of **R** where only states $|41\rangle$ and $|42\rangle$ ($|4,4\rangle$ and $|2,7\rangle$) have any significant presence. When the gate atom is moved to the extreme side of the ellipse (Fig. 2, fourth row), $|41\rangle$ no longer contributes but is replaced by a "whispering gallery" mode, $|43\rangle$ ($|0,10\rangle$). Here, an atom gates superpositions of $|42\rangle$ and $|43\rangle$, defining a new Hilbert 2-space and a new angle $\theta'$. The dual Hilbert spaces (Fig. 3a) comprise an overall 3-dimensional space; the crossover between planes in this space can be inferred from the structure of the wavefunctions (Fig. 3b-d).

To generate arbitrary superpositions, we require a way to predict the mirage-projecting wavefunction for any atom position **R**. We accurately reproduced our data using degenerate perturbation theory, representing the gate atom with a delta function potential: $V = \alpha \delta(\mathbf{r} - \mathbf{R})$. These perturbation theory results are implicit in microscopic Green's function results[16, 18] in our engineered limit that the separation between relevant corral wavefunctions is much less than their linewidth. Physically, a single atom placed inside the corral causes the degenerate wavefunctions to reorganize themselves into new zeroth-order wavefunctions. One is the superposition having the greatest amplitude possible at the location of the adatom and projects the mirage; this state is "tagged" and imaged by spectral mapping. All other combinations possess a node at **r** = **R** and are therefore insensitive to the perturbation and ignored by our differential measurements. The energy shift of the perturbed mode is proportional to $\alpha$; we detect a shift of a few meV at most, far less than the width of the states, so that this wavefunction remains energetically degenerate. Theoretical and experimental results for the Hilbert space angle while translating the gate atom across the two adjacent 2-spaces are shown in Fig. 3e; their excellent agreement demonstrates the validity of this analysis. Remarkably, for both spaces, all physically distinguishable combinations of the basis states can be created by the atom, limited only by the discreteness of adatom positions allowed by the substrate lattice.

The superpositions for atoms in regions where more than two wavefunctions have significant amplitude cannot be described by a single angle and are more difficult to visualize. In general, the state vector composed of the coefficients of the seven degenerate $|\psi(\mathbf{r})\rangle$ lies on the unit 7-sphere. In Fig. 4a we show "$a$ maps," which graphically display the theoretical superposition coefficients produced by an atom at each point in the ellipse. The *j*th map is a plot of $|a_j(\mathbf{R})|^2$. An atom at a white point in the *j*th map will generate a quantum state that is purely $|\Psi(\mathbf{r})\rangle = |\psi_j(\mathbf{r})\rangle$. Sliding that atom to an adjacent white spot in the *k*th map along a path where no other states contravene will continuously rotate the state vector from $|\psi_j(\mathbf{r})\rangle$ to $|\psi_k(\mathbf{r})\rangle$.

We find that almost all two-dimensional subspaces of the full 7-dimensional manifold can be fully indexed by a single atom. In Fig. 4b-d we give three examples of subspaces of **R** that index two-dimensional Hilbert spaces. The subspaces in Figure 4b and 4c encompass the experimental points shown in Figure 3; Supplementary Video 2 shows theoretical results for atom paths that further explore these spaces. In each example, we plot the two *a* maps as intersecting surfaces. Black lines demarcate the sets of **R** where $a_1^2 + a_2^2 > 0.8$ for all of the set and $\{|a_1|, |a_2|\} > 0.1$ for some point in each set—that is, where two and only two states are observable. This demonstrates a simple means to collapse a multidimensional Hilbert space onto a 2-dimensional subspace, within which a state vector can be rotated via real-space atomic gate manipulation.

Our methods have direct applicability to other lower-dimensional nanostructures. The potential imposed on the confined states of our quantum corral by the control atom is analogous to any sufficiently local and moveable gate potential. Selective gating of quantum superpositions in systems such as semiconductor[27-29] quantum dots should be readily achievable with scanning or otherwise mobile gates. The atom-gating method we have used also has analogies in cavity perturbation techniques in microwave resonators[30, 31]. Since quantum two-state systems can in general be mapped to a pseudospin, here the real-space atom position maps to the direction of an effective magnetic field that couples to this spinor. We anticipate that atomic gating will allow detection and manipulation of other fundamental phases, including the Berry and Aharonov-Bohm phases, by using atom paths traversing properly tuned quantum manifolds.

**Methods**

We used particle-in-an-elliptic-box wavefunctions $\psi_j$ to calculate the local density of states at position **r** and energy $\epsilon$,

$$LDOS(\mathbf{r}, \epsilon) = -\frac{1}{\pi} \mathrm{Im} \left( \sum_j \frac{|\psi_j(\mathbf{r})|^2}{\epsilon - \epsilon_j + i\delta_j} \right),$$

which we numerically integrated to recreate the expected topograph height[14],

$$z(\mathbf{r}) \propto \int_{E_\mathrm{F}}^{E_\mathrm{F}+V} LDOS(\mathbf{r},\epsilon)\, d\epsilon,$$

at the experimental sample bias of $V = 8$ meV.

The linewidth of the electron states has been alternately ascribed to tunnelling across the corral barrier[18, 24] or coupling to the bulk states through inelastic scattering at the walls[21] and intrinsic lifetime effects (e.g. electron-phonon interactions)[32]. Regardless of its origin, the broadening can be accounted for by a phenomenological lifetime added to particle-in-a-box wavefunctions. We used an electron self-energy $\delta = \Gamma/2 = 20\,\mathrm{meV}$ for all states, per the 40 meV linewidth $\Gamma$ observed in $dI/dV$ spectra for modes close to $E_\mathrm{F}$. Since contributions from these modes dominate the density of states relevant to our experiments, the variation of $\delta$ for modes far from $E_\mathrm{F}$ does not affect our calculation.

**Supplementary Information** is linked to the online version of the paper at www.nature.com/nature.

**Acknowledgements** This work was supported by the U.S. Office of Naval Research, the U.S. National Science Foundation, the U.S. Department of Energy, the Research Corporation, and the Stanford-IBM Center for Probing the Nanoscale. We acknowledge the U.S. NDSEG program (C.R.M.) and the Alfred P. Sloan Foundation (H.C.M) for fellowship support during this project. We thank D.-H. Lee, B. Sundaram, A. Bernevig, D. Haldane, and D. Eigler for discussions, and W. Mar for technical assistance. Correspondence and requests for materials should be addressed to H.C.M.

**Author Information** Reprints and permissions information is available at npg.nature.com/reprintsandpermissions. The authors declare no competing interests.

**References**

1. Crommie, M. F., Lutz, C. P. & Eigler, D. M. Confinement of electrons to quantum corrals on a metal surface. *Science* **262**, 218-220 (1993).

2. Stroscio, J. A. & Eigler, D. M. Atomic and molecular manipulation with the scanning tunneling microscope. *Science* **254**, 1319-1326 (1991).

3. Manoharan, H. C., Lutz, C. P. & Eigler, D. M. Quantum mirages formed by coherent projection of electronic structure. *Nature* **403**, 512-515 (2000).

4. Kouwenhoven, L. & Marcus, C. Quantum dots. *Physics World* **11**, 35-39 (1998).

5. Tans, S. J. *et al.* Individual single-wall carbon nanotubes as quantum wires. *Nature* **386**, 474-477 (1997).

6. Woodside, M. T. & McEuen, P. L. Scanned probe imaging of single-electron charge states in nanotube quantum dots. *Science* **296**, 1098-1101 (2002).

7. Eigler, D. M. *et al.* Information transport and computation in nanometre-scale structures. *Philosophical Transactions of the Royal Society of London A* **362**, 1135-1147 (2004).

8. Braun, K. F. & Rieder, K. H. Engineering electronic lifetimes in artificial atomic structures. *Physical Review Letters* **88**, 096801 (2002).

9. Rossi, E. & Morr, D. K. Spatially dependent Kondo effect in quantum corrals. *Physical Review Letters* **97**, 236602-236604 (2006).

10. Walls, J. D. & Heller, E. J. Spin-orbit coupling induced interference in quantum corrals. *Nano Letters* **7**, 3377-3382 (2007).

11. Duan, L. M., Cirac, J. I. & Zoller, P. Geometric manipulation of trapped ions for quantum computation. *Science* **292**, 1695-1697 (2001).

12. Zanardi, P. & Lloyd, S. Topological protection and quantum noiseless subsystems. *Physical Review Letters* **90**, 067902 (2003).

13. Crommie, M. F., Lutz, C. P. & Eigler, D. M. Imaging standing waves in a two-dimensional electron gas. *Nature* **363**, 524-527 (1993).

14. Tersoff, J. & Hamann, D. R. Theory of the scanning tunneling microscope. *Physical Review B* **31**, 805-813 (1985).

15. Aligia, A. A. Many-body theory of the quantum mirage. *Physical Review B* **64**, 121102 (2001).

16. Porras, D., Fernández-Rossier, J. & Tejedor, C. Microscopic theory for quantum mirages in quantum corrals. *Physical Review B* **63**, 155406 (2001).

17. Schmid, M. & Kampf, A. P. Mirages, anti-mirages, and further surprises in quantum corrals with non-magnetic impurities. *Annalen der Physik (Leipzig)* **12**, 463-470 (2003).

18. Aligia, A. A. & Lobos, A. M. Mirages and many-body effects in quantum corrals. *Journal of Physics: Condensed Matter* **17**, S1095-S1122 (2005).

19. Heller, E. J., Crommie, M. F., Lutz, C. P. & Eigler, D. M. Scattering and absorption of surface electron waves in quantum corrals. *Nature* **369**, 464-466 (1994).

20. Fiete, G. A. *et al.* Scattering theory of Kondo mirages and observation of single Kondo atom phase shift. *Physical Review Letters* **86**, 2392-2395 (2001).

21. Fiete, G. A. & Heller, E. J. Colloquium: Theory of quantum corrals and quantum mirages. *Reviews of Modern Physics* **75**, 933-948 (2003).

22. Agam, O. & Schiller, A. Projecting the Kondo effect: theory of the quantum mirage. *Physical Review Letters* **86**, 484 (2001).

23. Correa, A., Hallberg, K. & Balseiro, C. A. Mirages and enhanced magnetic interactions in quantum corrals. *Europhysics Letters* **58**, 899 (2002).

24. Rahachou, A. I. & Zozoulenko, I. V. Elastic scattering of surface electron waves in quantum corrals: Importance of the shape of the adatom potential. *Physical Review B* **70**, 233409 (2004).

25. Knorr, N., Schneider, M. A., Diekhoner, L., Wahl, P. & Kern, K. Kondo effect of single Co adatoms on Cu surfaces. *Physical Review Letters* **88**, 096804 (2002).

26. Lin, C. Y., Castro Neto, A. H. & Jones, B. A. Microscopic theory of the single impurity surface Kondo resonance. *Physical Review B* **71**, 35417 (2005).

27. Eriksson, M. A. *et al.* Cryogenic scanning probe characterization of semiconductor nanostructures. *Applied Physics Letters* **69**, 671-673 (1996).

28. Topinka, M. A. *et al.* Imaging coherent electron flow from a quantum point contact. *Science* **289**, 2323-2326 (2000).

29. Fallahi, P. *et al.* Imaging a single-electron quantum dot. *Nano Letters* **5**, 223-226 (2005).

30. Kuhl, U., Persson, E., Barth, M. & Stöckmann, H.-J. Mixing of wavefunctions in rectangular microwave billiards. *The European Physical Journal B-Condensed Matter* **17**, 253-259 (2000).

31. Gokirmak, A., Wu, D.-H., Bridgewater, J. S. A. & Anlage, S. M. Scanned perturbation technique for imaging electromagnetic standing wave patterns of microwave cavities. *Review of Scientific Instruments* **69**, 3410-3417 (1998).

32. Crampin, S. Electron states in quantum corrals. *Philosophical Transactions of the Royal Society of London A* **362**, 1149-1161 (2004).

**Figure 1 | Designed degeneracy in a quantum corral. a,** Eigenspectrum of desirable "double-peaked" modes of the elliptical box calculated analytically as a function of its deformation parameter $\mu = a/b$, where *a* and *b* are the semi-major and semi-minor axes of the ellipse, respectively. The spectrum is shown in terms of the dimensionless energy parameter $k^2 A$, where *A* is the area of the ellipse and $k = \sqrt{2m^* E/\hbar^2}$ is the wavenumber of an electron with effective mass $m^*$ and energy *E*. Intersecting dashed lines mark the parameters for our engineered degeneracy. **b,** Energy spectrum for a $\mu$ = 1.43 elliptical resonator with *a* = 79.2 Å (corresponding to circled degeneracy in **a**). **c,** 175 x 135 Å constant-*I* topograph (*V* = 8 mV, *I* = 1 nA) of the actual corral constructed. **d,** Simulated topograph calculated from the eigenmodes using a Green's function method. **e left,** Relative weights of each squared wavefunction composing the topograph in **d**, with corresponding $\psi(\mathbf{r})$ maps. **e right,** Energies of the seven modes in the corral lying closest to $E_F$. Schematic Lorentzians indicate the 40-meV observed linewidth of the states in *dI / dV* spectra. Vertical dashed lines mark the energy window for tunnelling electrons in our measurements. The Kondo resonance at $E_F$ with width $T_K$ = 53 K is shown shaded in red.

**Figure 2 | Single-atom gating and readout of quantum state superpositions. a-d**, Constant-*I* topographs (*V* = 8 mV, *I* = 1 nA) of the elliptical electron resonator containing a single control atom at four different locations $\mathbf{R}$. **e-h,** Simultaneously acquired *dI / dV* difference maps ($V_{ac}$ = 2 mV rms at 1007 Hz) manifesting the quantum mirage. A dotted line traces the locations of the wall atoms that have cancelled out. Each mirage reveals a specific superposition $\Psi(\mathbf{R})$, **i-l,** of two empty-corral eigenmodes that we control with the atomic gate. The internal atom thus acts as a handle for a single quantum phase angle in a two-dimensional Hilbert space, as indicated in **m-p**. These extracted phase angles are used to construct the superpositions graphed in **i-l.**

**Figure 3 | Complete indexing of two-dimensional Hilbert spaces. a**, Schematic of the two Hilbert space angles $\theta$ and $\theta'$. **b-d**, Close-up views (see black boxes in (**a**) of the three eigenstates superposed in Fig. 2: |4,4⟩, |2,7⟩, and |0,10⟩ (states $|41\rangle$, $|42\rangle$, and $|43\rangle$ respectively). The atom positions $\mathbf{R}$ in Fig. 2 are marked by crosses. **e,** Calculated Hilbert space phase angle as the control atom is translated along the major axis of the ellipse. Data points are the phase angles measured from Fig. 2m-p. The error bars indicate the accuracy limit of the fits of the difference maps (Fig. 2e-h) to the theoretical superpositions (Fig. 2i-l). Wavefunction amplitudes plotted with same colour scale as in Fig. 1e.

**Figure 4 | Superposition coefficient maps. a,** Squared superposition coefficients $a_j^* a_j$ of each numbered eigenmode *j* as a function of atom location $\mathbf{R}$. A white point in an eigenmode's *a*-map corresponds to an adatom position that projects a quantum mirage composed purely of that state; a black point means the state will not be observed. **b-d,** Examples of subsets of $\mathbf{R}$ indexing two-dimensional Hilbert spaces: $|41\rangle \leftrightarrow |42\rangle$, $|42\rangle \leftrightarrow |43\rangle$, and $|43\rangle \leftrightarrow |44\rangle$. The *a*-maps for two modes are superimposed as intersecting surfaces. Black lines delimit areas of the ellipse where no other modes have significant amplitude; a gate atom within these areas fully indexes a Hilbert space describable by a single phase angle $\theta$.

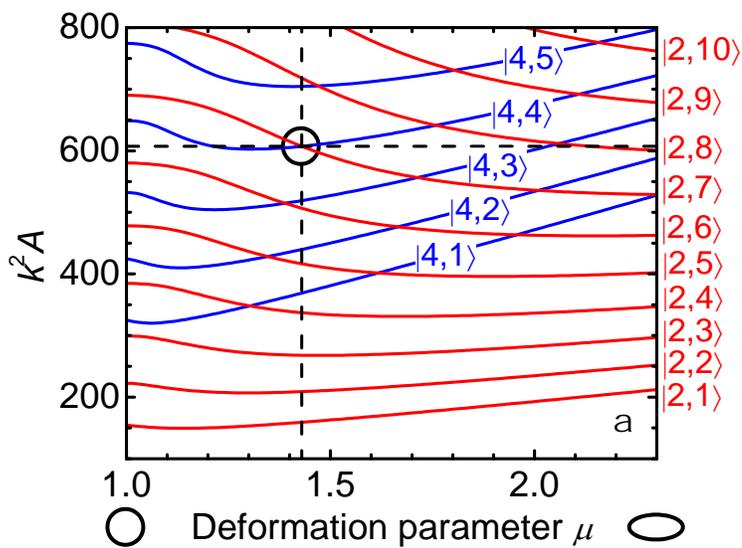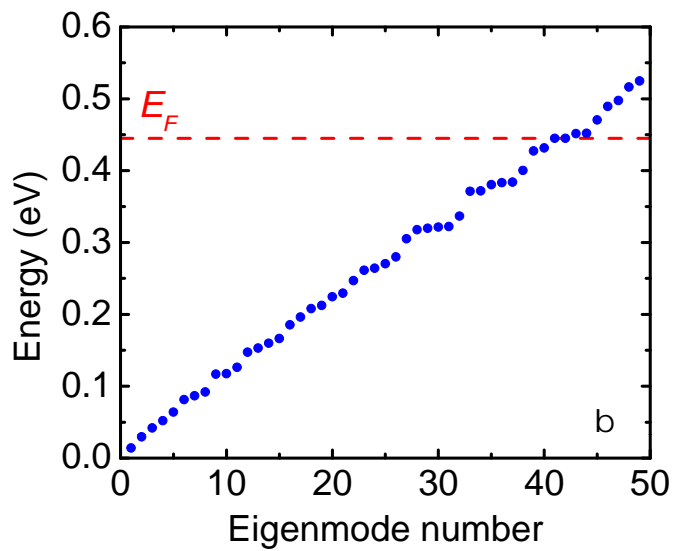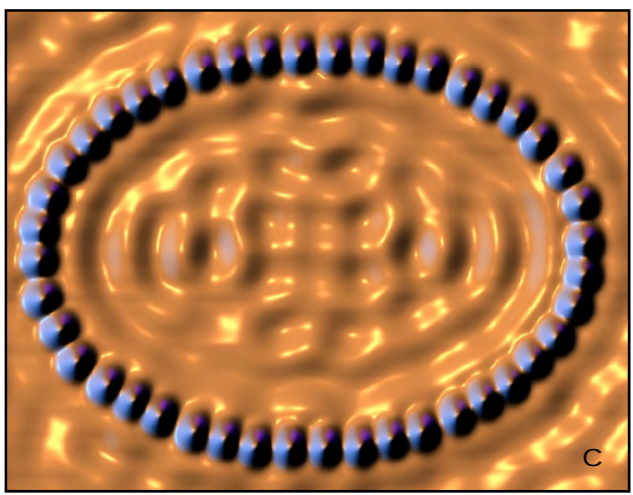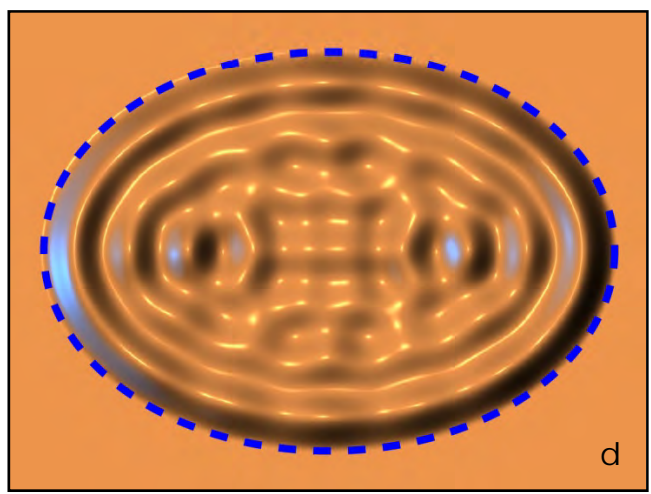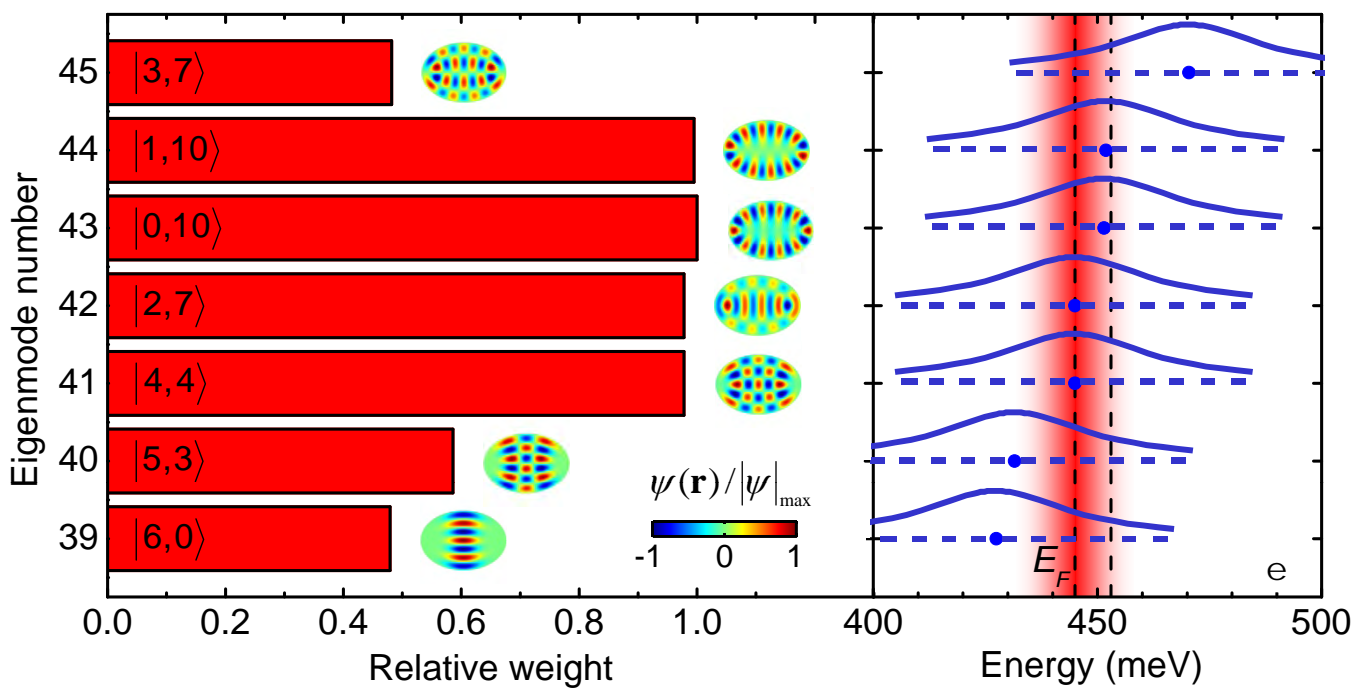

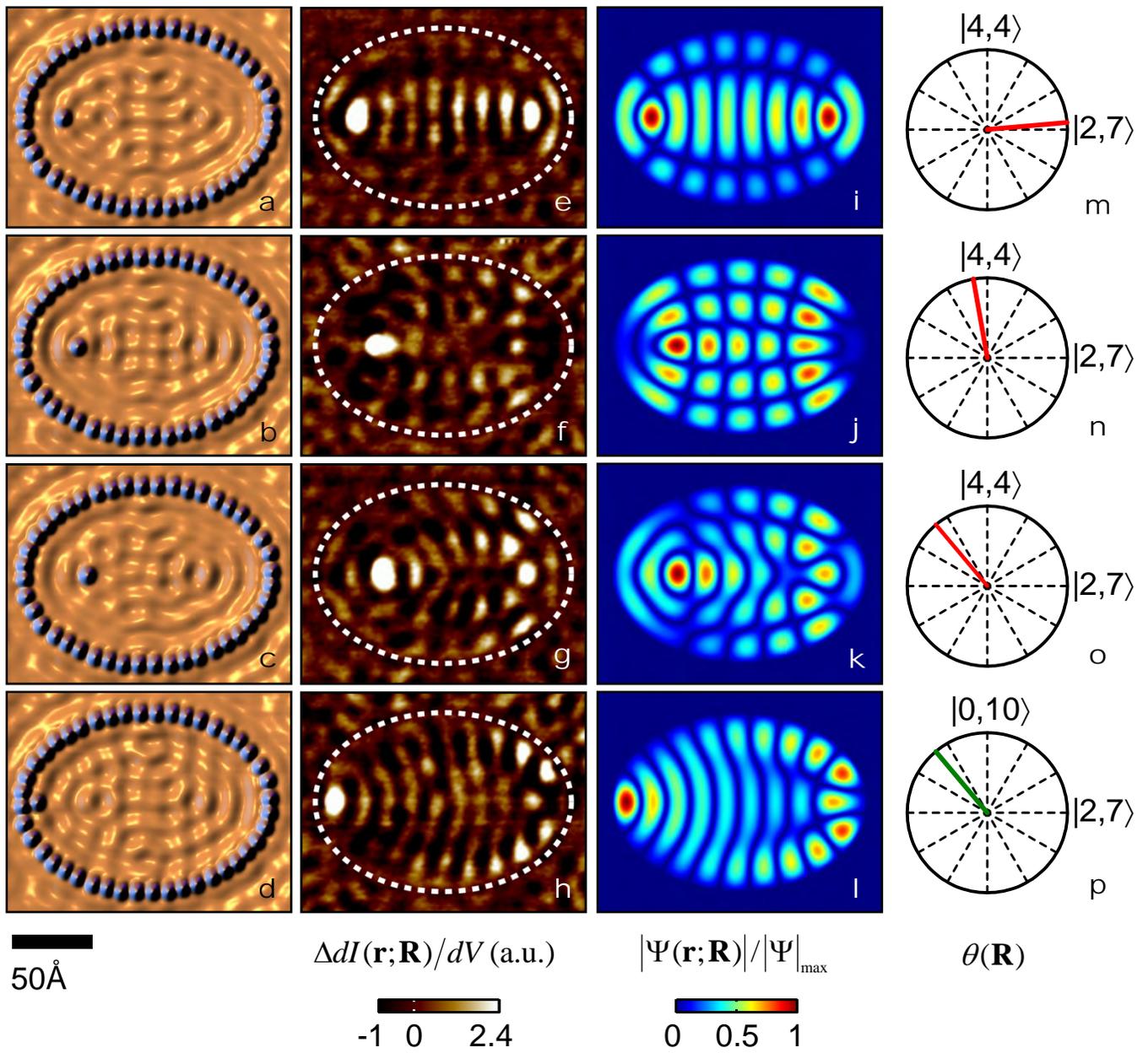

$\Delta dI(\mathbf{r};\mathbf{R})/dV$ (a.u.)    $|\Psi(\mathbf{r};\mathbf{R})|/|\Psi|_{max}$    $\theta(\mathbf{R})$

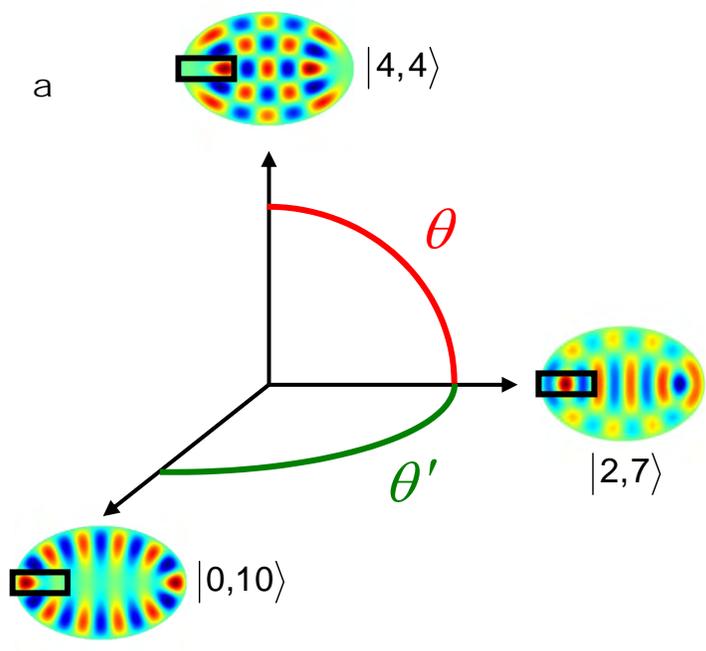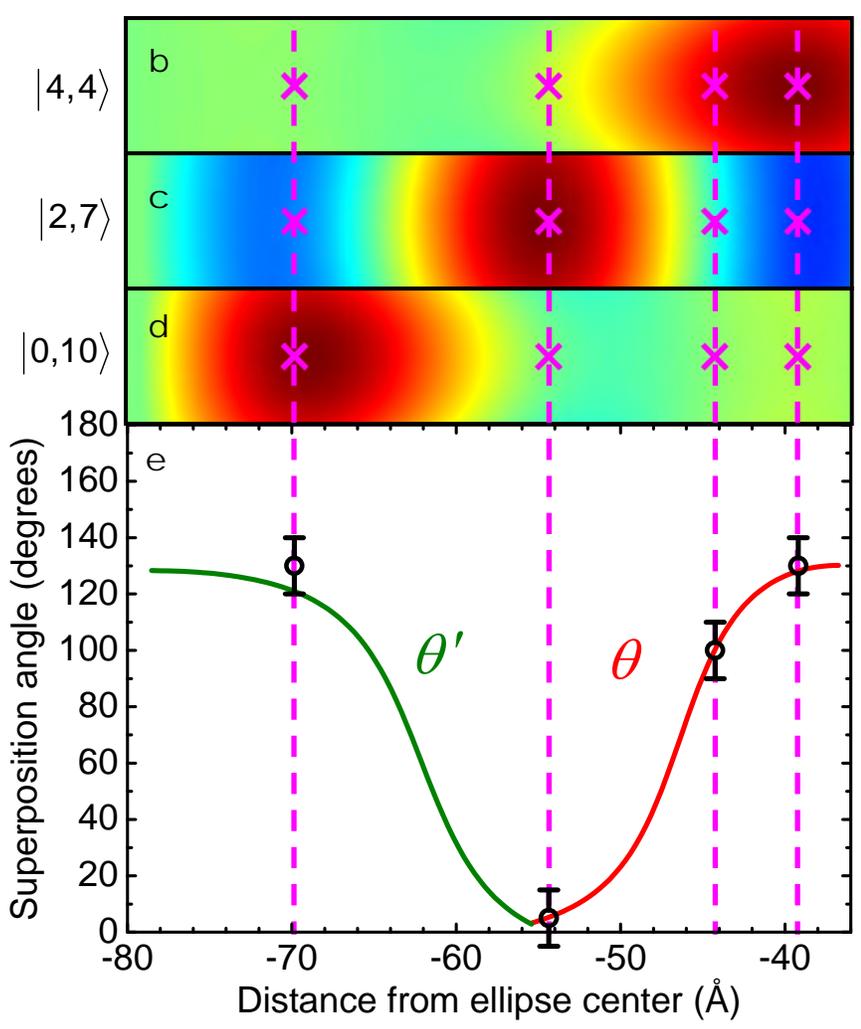

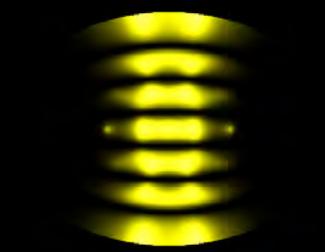
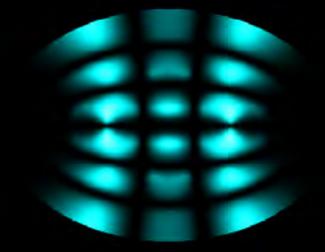
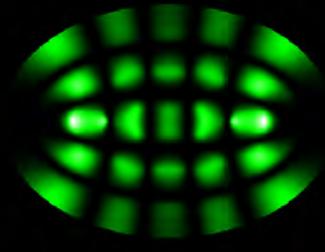
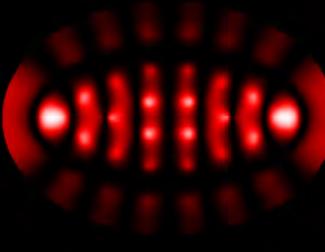
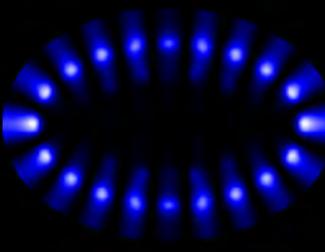
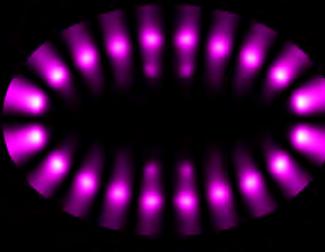
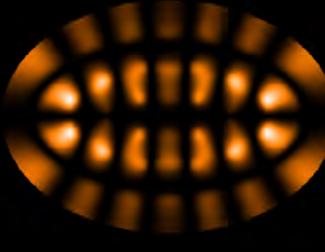
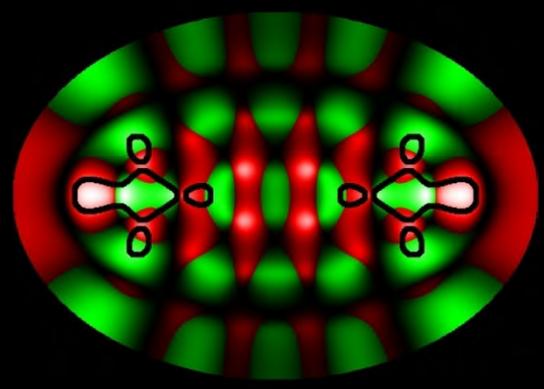
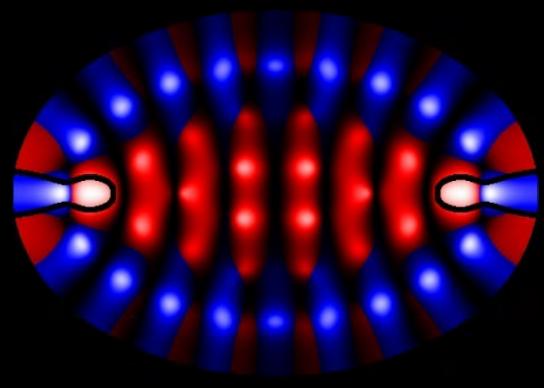
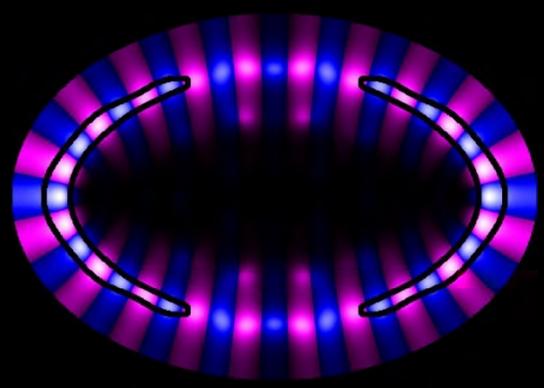
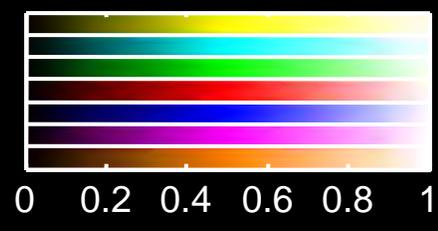

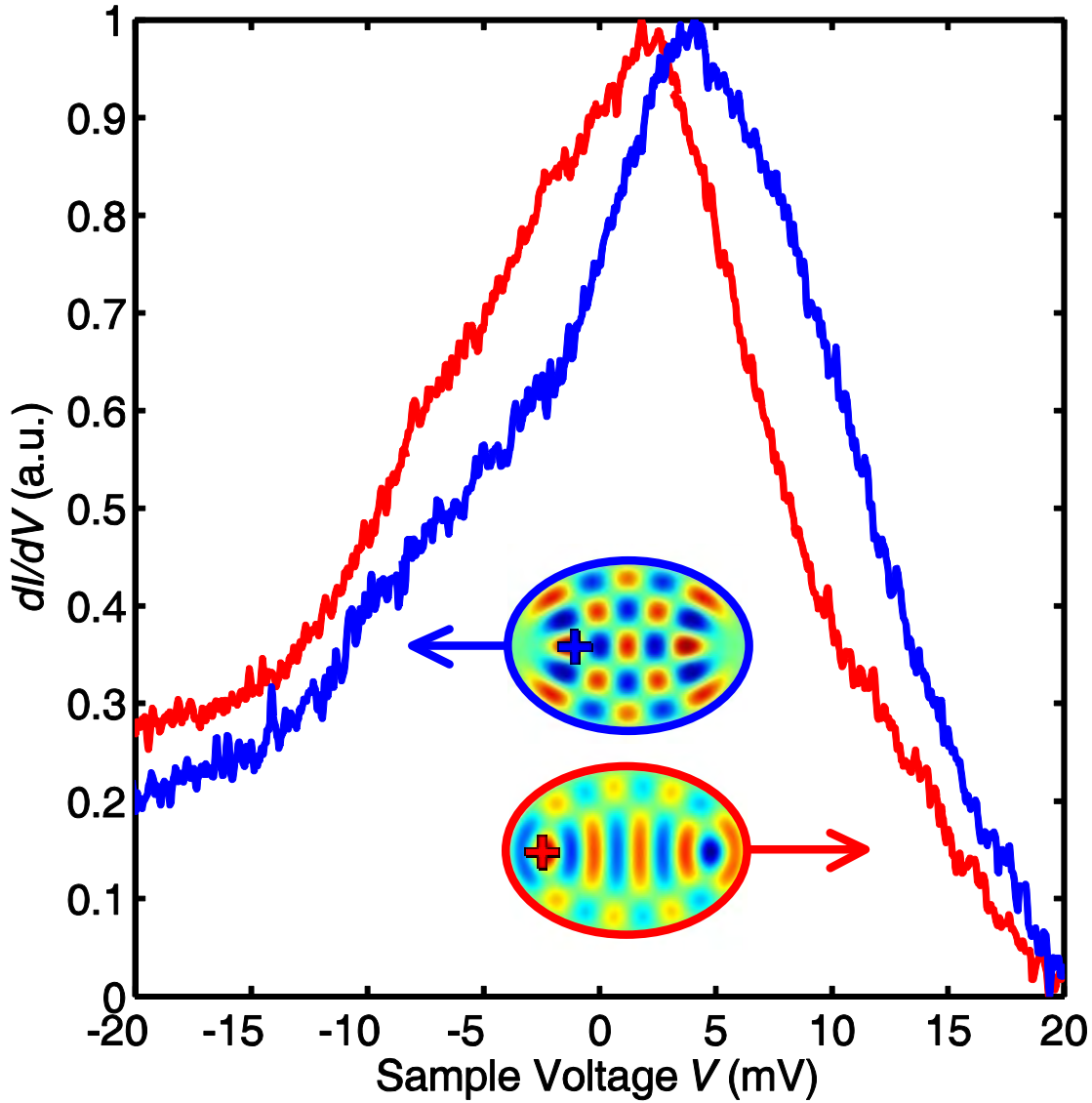

**Supplementary Figure 1 | Spectroscopy.** To verify that modes $|41\rangle$ and $|42\rangle$ are energetically degenerate, we performed $dI/dV$ vs. $V$ spectroscopy at real-space positions where each one is experimentally and theoretically greatest. The strongly peaked density of states at each point reveals that the two states are degenerate to within 2.5 mV. Note that the Co wall atoms become unstable at voltages $|V| \geq 20$ mV, limiting spectroscopy to this energy window.

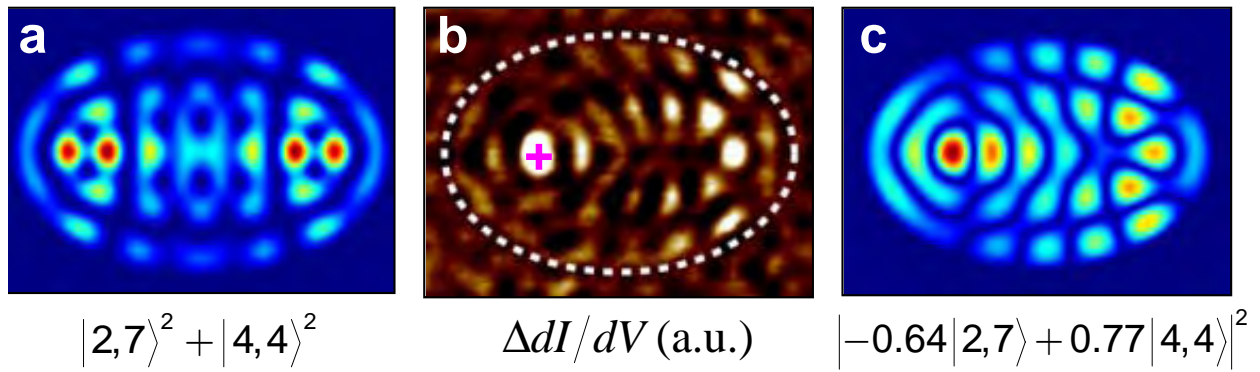

| a | b | c |
|---|---|---|
| $\lvert 2,7\rangle^{2}+\lvert 4,4\rangle^{2}$ | $\Delta dI/dV$ (a.u.) | $\lvert -0.64\lvert 2,7\rangle+0.77\lvert 4,4\rangle\rvert^{2}$ |

**Supplementary Figure 2 | Imaging superpositions.** **a**, By symmetry, squared eigenfunctions $\psi^2$ of an ellipse must be invariant under reflections across its major and minor axes. The same holds true for any combination $\sum_n c_n \lvert\psi_n\rangle^2$. For example, here the sum $\psi_{41}^{2}+\psi_{42}^{2}$ is plotted (see main text for definitions). **b**, When an atomic perturbation is introduced into the corral (pink cross), the symmetry of the system is broken and the response of the electrons (which we measure as a change in the differential conductance map) reflects new coherent superpositions $\left|\sum_n a_n \lvert\psi_n\rangle\right|^2$ (see Fig. 2 in main text). **c**, The coefficients $a_n$ are set by the position of the atom; here they are cos(130°) and sin(130°).

*Supplementary Video Summaries*

**Supplementary Video 1**

Superpositions as a function of phase angle.

(QuickTime; 7.4 MB).

*[ filename: superpositions_42_41.mov ]*

**Supplementary Video 2**

Selected real-space atom paths and their effects in state space.

(QuickTime; 4.7 MB).

*[ filename: twospaces.mov ]*